\documentclass[10pt,aps,prd,twocolumn,superscriptaddress,nofootinbib]{revtex4-1}
\usepackage{graphicx}
\usepackage{bigints}
\usepackage[normalem]{ulem}
\usepackage{cancel}
\usepackage{mathtools}
\usepackage{verbatim}
\usepackage{slashed}
\usepackage{titlesec}
\usepackage[usenames]{xcolor}
\usepackage{amsmath,amssymb}
\usepackage{mathrsfs}
\usepackage{amsthm}
\usepackage{hyperref}
\usepackage{appendix}
\usepackage{array}

\begin{document}

\title{Galilean fermions: Classical and quantum aspects}
\author{Aditya Sharma}
\email{p20170442@goa.bits-pilani.ac.in, \\
adityasharma.theory@gmail.com}
\affiliation{BITS-Pilani, KK Birla Goa Campus, NH 17B, Bypass Road, Zuarinagar, Goa, India 403726}
\begin{abstract}
We study the classical and quantum ``properties" of Galilean fermions in $3+1$ dimensions. We have taken the case of massless Galilean fermions minimally coupled to the scalar field. At the classical level, the Lagrangian is obtained by null reducing the relativistic theory in one higher dimension. The resulting theory is found to be invariant under infinite Galilean conformal symmetries. Using Noether's procedure, we construct the corresponding infinite conserved charges. Path integral techniques are then employed to probe the quantum ``properties" of the theory. The theory is found to be renormalizable. A novel feature of the theory is the emergence of \emph{mass} scale at the first order of quantum correction. The conformal symmetry of the theory breaks at quantum level. We confirm this by constructing the beta function of the theory.	`
\end{abstract}
\maketitle

\section{Introduction}

Symmetries are essential in the study of any physical system, in that they are responsible for conservation laws. For example, symmetry in space and time translation leads to momentum and energy conservation laws respectively. It is well established that Lorentz symmetry is essential to describe the physics of fundamental particles and their interactions. However, for systems where the speed of objects $(v)$ involved is much less than the speed of light $(c)$ (i.e, $v<<c$, also known as non-relativistic limit), Galilean symmetry is better suited. The emergence of non relativistic symmetries in the study of cold atoms, fermi condensates and Efimov effect etc \cite{Nishida:2010tm}\cite{Bedaque:1998km}\cite{Hartnoll:2009ns} has further fuelled the validity to consider non relativistic limits.\\[4pt]
Recently, there has been an upsurge to construct field theories consistent with Galilean symmetry. This is because Galilean symmetry has paved its way in describing condensed matter systems such as quantum hall effect, non-relativistic fluid dynamics and magnetohydrodynamics \cite{2013arXiv1306.0638T}\cite{2000physics..10042J}\cite{1998hep.th....9123J}. Galilean symmetry is characterized by unequal scaling of space and time (also known as Galilean limit\footnote{The Galilean limit is same as non-relativistic limit. They are often used interchangeably for each other in the literature.}) i.e
\begin{equation*}
 t \to t\;\;,\;\; x_i \to \epsilon x_i \;\;, \;\; \epsilon \to 0. 
 \end{equation*}
and is described by a set of symmetry generators viz. spatial and temporal translations $(P_i, H)$, homogeneous spatial rotations $(J_{ij})$ and Galilean boosts $(B_i)$. In addition, Galilean symmetry can also be conformally extended by including spatial conformal transformations $(K_i)$, temporal conformal transformations $(K)$ and dilatations $(D)$. Together, they form a closed Lie algebra known as finite Galilean conformal algebra (fGCA) \cite{Bagchi:2017yvj}\cite{Bagchi:2014ysa}. The Galilean conformal symmetry generators can be obtained either, by taking Galilean limit of Poincar\'{e} symmetry generators \cite{Bagchi:2014ysa} or, by finding the conformal isometries of Newton-Cartan manifolds (a brief discussion is given in subsection \ref{section:gca}. For more details see \cite{2009}\cite{Duval_1993}).  A remarkable feature of fGCA is that it can be given an infinite lift to construct an infinite Lie algebra \eqref{eqn:lift} known as infinite Galilean conformal algebra (GCA) \cite{Bagchi:2017yvj}\cite{Bagchi:2014ysa}\cite{Bagchi:2015qcw}\cite{Bagchi:2022twx}.\\[5pt]
The study of Galilean conformal field theories has recently seen a revival \cite{Bagchi:2017yvj}\cite{Bagchi:2014ysa}\cite{Bagchi:2015qcw}\cite{Bagchi:2022twx}\cite{Banerjee:2022eaj}. This is mainly because field theories consistent with GCA admits infinite number of conserved charges at classical level (see \cite{Bagchi:2015qcw}\cite{Bagchi:2022twx} and references therein). Surprisingly, not much heed has been payed to understand the quantum ``properties" of these conformal theories except for some recent work in  \cite{Banerjee:2022uqj}. Addressing the issue of quantization of Galilean conformal field theories is also important because of its application in many physics systems \cite{Lukierski:2005xy}\cite{Son:2005rv}\cite{Jackiw:2000tz}. In this paper we present both, the classical and quantum field description of massless Galilean fermions minimally coupled to Galilean scalar in $3+1$-dimensions. Owing to the interaction between a scalar field and fermionic field, we call the resultant theory Galilean Yukawa theory.\\[5pt]
Some of the early work on Galilean fermion was carried out by L\'{e}vy-Leblond in 1967 \cite{Levy-Leblond1967} where a Galilei invariant analogue of Dirac equation was constructed. An interesting finding was that the spin magnetic moment, with its Land\'{e} factor $g=2$, is not a relativistic property. Galilean fermions have also been studied in \cite{Duval:1995fa} where a massless Dirac equation was shown to exhibit the Schr\"{o}dinger symmetry (which we know is a conformal extension of Galilean group \cite{2009}). Recent investigation on Galilean fermion are considered in \cite{Banerjee:2022uqj}\cite{Hao:2022xhq}\cite{Yu:2022bcp}. The Galilean Yukawa theory is the simplest example of an interacting conformal field theory admitting fermionic degrees of freedom, consistent with GCA. The theory becomes even more captivating at quantum level because the \emph{mass} term surfaces at the first order of quantum correction. Admittance of the \emph{mass} scale in a pure Galilean field theory upon renormalization has never been addressed before in the literature. \\[5pt]
Recent study carried out with Galilean quantum electrodynamics in $(3+1)$- dimensions \cite{Banerjee:2022uqj} suggests the presence of global conformal anomalies in the theory at quantum level which is quite different to the case of Galilean electrodynamics coupled to Schr\"{o}dinger scalar (sGED) in $(2+1)$-dimensions \cite{Chapman:2020vtn} where the beta function vanishes identically leading to a family of non relativistic conformal fixed points. $\mathcal{N}=2$ supersymmetric extension of Galilean Electrodynamics in $(2+1)$ dimensions constructed has also been studied in \cite{2022arXiv220706435B}. 
It must be noted that the free Galilean scalar field theory does not admit any dynamical degrees of freedom. This is because the Galilean limit kills the kinetic part of the theory. Coupling Galilean fermions to the Galilean scalar field introduces the dynamical degrees of freedom into the theory. This makes for an interesting case of a Galilean invariant conformal field theory that admits infinite number of conserved charges at the classical level. Thus, this paper is an attempt to present both, the classical and quantum field description of Galilean Yukawa theory.\\[5pt]
The Galilean Yukawa theory is constructed in this paper by null reducing the relativistic Yukawa theory in one higher dimensions. This method is well known in the literature and goes by the name of ``null reduction" \cite{1995}\cite{Santos_2004}\cite{Bergshoeff_2016}\cite{20.500.11850/488630}. At the classical level, the theory admits an infinite number of conserved charges. In order to describe the quantum field description, we employ path integral techniques. The method of cut-off regularization has been employed to regulate the UV divergences and the theory is then renormalized upto 1 loop. Interestingly, the \emph{mass} scale in the scalar sector appears at first order of quantum correction. The admittance of \emph{mass} term assures that the theory is no longer scale invariant at the quantum scale, albeit exhibiting conformal invariance at the classical level. This is suggestive of global conformal anomaly in the theory\cite{Jensen:2014hqa}\cite{Jain:2015jla}, which is further guaranteed by the non vanishing nature of the beta function. \\[5pt]
This paper is organized as follows: In section \ref{section:gca1} we present the classical field description of Galilean Yukawa theory. We briefly discuss the Galilean conformal symmetry in subsection \ref{section:gca} and present some of the well known results in the literature. In subsection \ref{section:lagrangian}, we construct the Lagrangian for Galilean Yukawa theory using null reduction. We address the symmetries of the theory in subsection \ref{section:scc} followed by the construction of conserved charges. We delve further into the theory in section \ref{section:quantum} by presenting the quantum field description of Galilean Yukawa theory. We have employed functional techniques to develop the quantum field description of the theory. To bring out the nature of divergences in the theory we employ the method of cut-off regularization. To this end, we evaluate the 1 loop corrections to the propagators and vertex in subsection \ref{section:regular}. The issue of renormalization is addressed in subsection \ref{section:renor} followed by the summary and discussions in section \ref{section:conclusion}.

 \section{Classical field description of Galilean Yukawa theory}
 \label{section:gca1}
 \subsection{Galilean Conformal Symmetry}
 \label{section:gca}
 Galilean conformal symmetry of a $(d+1)$-dimensional spacetime is described by the set of symmetry generators- time translations $(H)$, space translations $(P_i)$, homogeneous rotations $(J_{ij})$, Galilean boosts $(B_i)$, dilatation $(D)$ and spatial and temporal conformal transformations $(K_i, K)$. In an adapted coordinate system $x^\mu\equiv (t,x^i)$ we have
\begin{equation}
\label{eqn:galsym}
\begin{split}
&H=-\partial_t \;\;,\;\; P_i =\partial_i\;\;,\;\; J_{ij}= x_i \partial_j-x_j \partial_i\\
&B_i= t \partial_i\;\;\; D=-(t \partial_t+x^i\partial_i)\;\;,\;\;K_i= t^2\partial_i \\
&K=-(t^2\partial_t+2x_i t \partial_i)
\end{split}
\end{equation}
The symmetry generators \eqref{eqn:galsym} except $J_{ij}$ can be cast into a compact notation i.e,
\begin{eqnarray*}
\label{eqn:algebra}
&L^{(n)}=& -t^{n+1}\partial_t-(n+1)t^nx_i\partial_i \\
&M_{i}^{(n)}=&t^{n+1}\partial_i 
\end{eqnarray*}
where $H,D,K$ can be recovered by setting $n=-1,0,1$ in $L^{(n)}$ and $P_i,B_i,K_i$ are recovered by setting $n=-1,0,1$ in $M_{i}^{(n)}$. The generators $L^{(n)}, M_{i}^{(n)} , J_{ij}$ form a closed Lie algebra called finite Galilean conformal algebra (fGCA) given by
\begin{equation}
\begin{split}
\label{eqn:lift}
&\left[L^{(n)}, L^{(m)}\right]=(n-m) L^{(n+m)}\\ 
&\left[L^{(n)}, M_{i}^{(m)}\right]=(n-m) M_{i}^{(n+m)}\\ 
&\left[M_{i}^{(n)}, M_{j}^{(m)}\right]=0\\
&\left[L^{(n)}, J_{i j}\right]=0\\
&\left[J_{i j}, M_{k}^{(n)}\right]=M_{[j}^{(n)} \delta_{i] k}
\end{split}
\end{equation}
A striking feature of \eqref{eqn:lift} is that the algebra closes $\forall \;n,m \in \mathbb{Z}$. This gives fGCA an infinite lift. The resulting Lie algebra is called infinite Galilean conformal algebra (GCA). The reason for the infinite lift is captured in the underlining geometry. Precisely, the Galilean conformal symmetries are related to the conformal isometries of a `flat' Newton-Cartan spacetime (see \cite{2009}\cite{Duval_1993}\cite{PhysRevD.31.1841}\cite{2014}\cite{https://doi.org/10.48550/arxiv.2112.13403}). A Newton-Cartan (NC) spacetime is a $(d+1)$ dimensional smooth manifold equipped with a degenerate contravariant metric $g$ along with a non-vanishing 1-form $\theta$ which also happens to be in the Kernel of $g$. \\[5pt]
The conformal isometries of NC spacetime are those vector field $X$ that preserves $\theta$ and $g$ upto a non trivial conformal factor $\lambda$ \cite{2009}\cite{2014c} i.e
\begin{equation}
\label{eqn:isom}
\pounds_{X}g=\lambda g \;\;,\;\; \pounds_{X}\theta=-\frac{1}{2}\lambda \theta
\end{equation}
For a flat NC spacetime $(\mathbb{R}\times \mathbb{R}^d)$, in an adaptive coordinate chart $x^\mu \equiv(t,x^i)$
\begin{equation*}
g=g^{\mu \nu}\partial_\mu \otimes \partial_\nu \;\;,\;\; \theta=dt\;\;\; \text{where}\;\; g^{\mu \nu}= diag(0,I)
\end{equation*}
(\ref{eqn:isom}) reduces to
\begin{equation}
\label{eqn:vec}
X= \alpha(t) \frac{\partial}{\partial t}+\Big(\omega_{ij}(t)x^j+x^i \beta(t)+\xi^i(t)\Big)\frac{\partial}{\partial x^i}
\end{equation}
where  $\omega \in SO(d)$, $\alpha , \beta \in \mathbb{R}$ and $\xi \in \mathbb{R}^d$ are the arbitrary functions of time; explaining the infinite lift of GCA.\\[5pt]
The generators $\{L^{(n)}, M_{i}^{(n)}, J_{ij}\}$ can be used to construct the action of symmetry generators on the local fields. This can be done either by looking at scale-boost representations or scale-spin representations of GCA. In this paper we shall employ the scale-spin representation of GCA \cite{2018} (for scale-boost representations of GCA we request the reader to see \cite{Bagchi:2009ca}\cite{Bagchi:2009pe}). For some general field $\Phi=(\varphi, \phi, A_i,.....)$, where $\varphi$ is some scalar field, $\phi$ is a 2 component spinor, $A_i$ is a vector field and the dots represents higher spin fields, we can write,
\begin{eqnarray}
\label{eqn:l}
&\delta_{L^{(n)}} \Phi&= \Big(t^{n+1}\partial_t +(n+1) t^n (x^l \partial_l+\Delta) \Big) \Phi \nonumber \\
&\qquad& \qquad \qquad \qquad -t^{n-1} n(n+1)x^k \delta_{B_k} \Phi \\[5pt]
\label{eqn:m}
&\delta_{M_{l}^{(n)}} \Phi&= -t^{n+1} \partial_l \Phi+(n+1)t^n \delta_{B_l} \Phi \\[5pt]
\label{eqn:j}
&\delta_{J_{ij}} \Phi&= (x_i \partial_j -x_j \partial_i )\Phi + \Sigma_{ij} \Phi
\end{eqnarray}
where $\Delta$ is the scaling dimension, $ \delta_{B_l} \Phi$ is the action of boost generator on the field $\Phi$ and $\Sigma_{ij} =\frac{1}{4} [\sigma_i , \sigma_j ]= \frac{i}{2}\epsilon_{ijk} \sigma_k  $. For scalar field $\varphi$, $\Sigma_{ij} \varphi=0$. We will employ \eqref{eqn:l}-\eqref{eqn:j} to establish the invariance of Galilean Yukawa theory \eqref{eqn:complag} under GCA and later again to construct the conserved charges for the theory.

 \subsection{Lagrangian formulation}
 \label{section:lagrangian}
We shall employ the method of null reduction to construct the Lagrangian for the Galilean Yukawa theory. The method of null reduction has been widely used in the literature to construct the Lagrangians for non-relativistic field theories \cite{1995}\cite{Santos_2004}\cite{Bergshoeff_2016}\cite{20.500.11850/488630}.  We start with a relativistic theory in one higher dimension. In an adaptive coordinate chart $x^\mu=(u,t,x^i)$, where $u,t$ are the two real null coordinates and $x^i=(x,y,z)$ are the spatial coordinates, we write,
\begin{equation}
\label{eqn:lagrel}
\tilde{\mathcal{L}}=\frac{1}{2}\eta^{\mu \nu} (\partial_\mu \varphi)( \partial_\nu \varphi )+i \bar{\psi}\gamma^\mu \partial_\mu \psi -g \bar{\psi}\psi \varphi
\end{equation}
where $\eta^{\mu \nu}$ is the metric tensor for the Minkowski line element in the coordinate chart $x^\mu$ i.e, 
\begin{equation}
\label{eqn:nullmetric}
ds^2=du \otimes dt +dt \otimes du+\delta_{ij} dx^i \otimes dx^j=\eta_{\mu \nu} dx^\mu \otimes dx^\nu
\end{equation}
$\varphi$ is the scalar field, $\psi$ is a 4 component spinor, $g$ is the coupling strength and $\gamma^\mu$ are the Dirac matrices whose explicit form in coordinate chart $x^\mu$ is taken to be 
\begin{eqnarray}
\label{eqn:gamma}
\gamma^u= \begin{pmatrix}
0\;&\;\;\sqrt{2} \\
0\;&\;\;\;\;0\\
\end{pmatrix}\\
\gamma^t=\begin{pmatrix}
0\;&\;\;0 \\
-\sqrt{2}\;&\;\;0\\
\end{pmatrix} \\
\gamma^i= \begin{pmatrix}
i\sigma^i&\;\;0 \\
0&-i\sigma^i\\
\end{pmatrix}
\end{eqnarray}
where $\sigma^i$ are the usual Pauli matrices. The $\gamma$ matrices (\ref{eqn:gamma}) obey the standard Clifford algebra,
$\{ \gamma^\mu, \gamma^\nu \}= -2\eta^{\mu \nu}$
where $\eta^{\mu \nu}$ is the metric tensor associated to the Minkowski line element (\ref{eqn:nullmetric}). The $\gamma$-matrices allows us to define the adjoint of the $\psi$, i.e $\bar{\psi}=\psi^\dagger G$, where\footnote{for a step-by-step construction of $\gamma$-matrices, $G$ and $\bar{\psi}$ we request the reader to check our previous work \cite{Banerjee:2022uqj}} 
\begin{equation}
G =\begin{pmatrix}
\;0\;\;&\;\;1\;\\
\;1\;\;&\;\;0\;\\
\end{pmatrix}
\end{equation}
Now to write down the Lagrangian for the Galilean field theory we null reduce \eqref{eqn:lagrel} along the null direction $u$ i.e, we demand
\begin{equation*}
\partial_u \varphi=\partial_u \psi=0
\end{equation*}
This leads to the Lagrangian $\mathcal{L}$ for the Galilean Yukawa theory
\begin{equation}
\label{eqn:lagnonrel}
\mathcal{L}=\frac{1}{2}(\partial_i \varphi)( \partial_i \varphi )+i \bar{\psi}(\gamma^t \partial_t +\gamma^i \partial_i)\psi -g \bar{\psi}\psi \varphi
\end{equation}
Note that the `leftover' null coordinate $t$ has now acquired the status of the non-relativistic time. The gamma matrices for \eqref{eqn:lagnonrel} are now given by $\gamma^I=(\gamma^t,\gamma^i)$ and they obey the degenerate Clifford algebra given by
\begin{equation}
\{\gamma^I , \gamma^J\}=-2 g^{IJ}
\end{equation}
where $g^{IJ}$ is the degenerate metric i.e, $g^{IJ}=\text{diag}(0,1,1,1)$ on the Newton-Cartan spacetime. We can further reduce \eqref{eqn:lagnonrel} in terms of the components of $\psi$ i.e, we write $\psi =\begin{pmatrix}
\phi \\
\chi
\end{pmatrix}$
where, $\phi$ and $\chi$ are the 2 component spinors themselves, allowing us to write \eqref{eqn:lagnonrel} as
\begin{equation}
\begin{split}
\label{eqn:complag}
\mathcal{L}=\frac{1}{2}(\partial_i \varphi)( \partial_i \varphi )-\sqrt{2} i \phi^\dagger \partial_t \phi -\chi^\dagger \sigma^i \partial_i \phi \\
+\phi^\dagger \sigma^i \partial_i \chi-g\varphi(\chi^\dagger \phi+\phi^\dagger \chi)
\end{split}
\end{equation}
Using \eqref{eqn:complag}, we can write $\mathbb{L}=\bigintsss d^3 x \; \mathcal{L}$ i.e, 
 \begin{equation}
 \begin{split}
 \label{eqn:L}
\mathbb{L}=\bigintssss d^3 x \; \bigg\{ \frac{1}{2}(\partial_i \varphi)( \partial_i \varphi )-\sqrt{2} i \phi^\dagger \partial_t \phi -\chi^\dagger \sigma^i \partial_i \phi \\
+\phi^\dagger \sigma^i \partial_i \chi-g\varphi(\chi^\dagger \phi+\phi^\dagger \chi)\bigg\}
\end{split}
 \end{equation}
Note that in the absence of the spinor field, the theory reduces to that of a real Galilean scalar field which does not exhibit any dynamics. Thus, the coupling of spinor field can also be understood as the introduction of matter degree of freedom into the Galilean scalar field theory. Variation of the Lagrangian, results in the following equations of motion\footnote{For full spectrum of equations of motion, one needs to include the complex conjugates of \eqref{eqn:eomi}-\eqref{eqn:eomf}}
\begin{eqnarray}
\label{eqn:eomi}
&-\sqrt{2} i \partial_t \phi+\sigma^i \partial_i \chi-g \varphi \chi=0\\[5pt]
&\partial^2 \varphi +g (\chi^\dagger \phi+\phi^\dagger \chi)=0\\[5pt]
\label{eqn:eomf}
&\sigma^i\partial_i \phi+g \varphi \phi=0
\end{eqnarray}
For completeness, we mention the canonical\footnote{Note that the canonical momentum ($\pi_\varphi$) for the Galilean scalar $\varphi$ does not appear in (\ref{eqn:hamiltonian}). This is because $\pi_\varphi$ is a primary constraint in the theory and shall only appear in the expression for the total Hamiltonian. Total Hamiltonian becomes essential if we were to address quantization via canonical techniques. However, in this paper we are interested in path integral quantization. For more details on canonical quantization of systems with constraints see \cite{henneaux1992quantization}.} Hamiltonian $H$ for the Galilean Yukawa theory,
\begin{equation}
\begin{split}
\label{eqn:hamiltonian}
H= -\bigintsss d^3 x\; \bigg(\frac{1}{2}(\partial_i \varphi)( \partial_i \varphi )-\chi^\dagger \sigma^i \partial_i \phi +\phi^\dagger \sigma^i \partial_i \chi \\
-g\varphi(\chi^\dagger \phi+\phi^\dagger \chi)\bigg)
\end{split}
\end{equation}

\subsection{Symmetries and Conserved Charges}
\label{section:scc}
Let us now analyse the symmetries of Galilean Yukawa theory constructed above. In order to do that, we first have to write down the action of Galilean symmetry generators $\{L^{(n)}, M_{i}^{(n)}, J_{ij}\}$ on the fields $(\varphi, \phi, \chi)$. We employ \eqref{eqn:l}-\eqref{eqn:m} to achieve that. The action of $M_{i}^{(n)}$ on the fields takes on the following form
\begin{eqnarray}
\label{eqn:actionm1}
&\delta_{M_{i}^{(n)}} \phi&= -t^{n+1} \partial_i \phi\\[5pt]
&\delta_{M_{i}^{(n)}} \chi&= -t^{n+1} \partial_i \chi-\frac{i (n+1)}{\sqrt{2}} t^n \sigma_i \phi\\[5pt]
\label{eqn:actionm2}
&\delta_{M_{i}^{(n)}} \varphi&= -t^{n+1} \partial_i \varphi
\end{eqnarray}
 and the action of $L^{(n)}$ on the fields read
 \begin{eqnarray}
 \label{eqn:actionl1}
 &\delta_{L^{(n)}} \phi&= t^{n+1} \partial_t \phi+(n+1)t^n (x^j\partial_j+\Delta_1)\phi\\[5pt]
 &\delta_{L^{(n)}} \chi&= t^{n+1} \partial_t \chi+(n+1)t^n (x^j\partial_j+\Delta_1)\chi \nonumber \\
 &\qquad& \qquad \qquad +\frac{i}{\sqrt{2}}n(n+1)t^{n-1}x^k\sigma_k \phi\\[5pt]
 \label{eqn:actionl2}
 &\delta_{L^{(n)}} \varphi&= t^{n+1} \partial_t \varphi+(n+1)t^n (x^j\partial_j+\Delta_2)\varphi
 \end{eqnarray}
 where $\Delta_1=3/2$ and $\Delta_2=1$. The stage is now set to address the symmetries of Galilean Yukawa theory. We shall address the symmetries from the Lagrangian perspective. 
 \subsubsection{Symmetries of Lagrangian}
 We begin by varying the Lagrangian of the theory \eqref{eqn:L} by an arbitrary variation $\delta$ i.e,
 \begin{equation}
 \label{eqn:variation}
 \begin{split}
 \delta \mathbb{L}=  \bigintssss d^3 x \; \bigg\{ (\partial_i \varphi)(\partial_i \delta \varphi)-\sqrt{2} i \delta \phi^\dagger \partial_t \phi-\sqrt{2} i \phi^\dagger (\partial_t \delta \phi)\\
 -\delta \chi^\dagger \sigma^i \partial_i \phi-\chi^\dagger \sigma^i \partial_i\delta\phi+\delta \phi^\dagger \sigma^i \partial_i \chi +\phi^\dagger \sigma^i \partial_i \delta \chi\\
 -g \delta \varphi \big(\chi^\dagger \phi+\phi^\dagger \chi\big)-g\varphi \big(\delta \chi^\dagger \phi +\chi^\dagger \delta \phi+\phi^\dagger \delta \chi+\delta \phi^\dagger \chi\big)\bigg\}
 \end{split}
 \end{equation}
To arrive at the symmetries of the Lagrangian under Galilean conformal generators, we restrict $\delta$ to $\delta_{M_{i}^{(n)}}$ and $\delta_{L^{(n)}}$. Now upon using \eqref{eqn:actionm1}-\eqref{eqn:actionm2} we arrive at
\begin{equation}
\delta_{M_{i}^{(n)}} \mathbb{L}= 0
\end{equation}
Also, using \eqref{eqn:actionl1}-(\ref{eqn:actionl2}), we can end up with
\begin{equation}
\begin{split}
\delta_{L^{(n)}} \mathbb{L}= \bigintsss d^3x \; \partial_t \bigg( \frac{1}{2}(\partial_i \varphi)( \partial_i \varphi )-\sqrt{2} i \phi^\dagger \partial_t \phi -\chi^\dagger \sigma^i \partial_i \phi \\
+\phi^\dagger \sigma^i \partial_i \chi-g\varphi(\chi^\dagger \phi+\phi^\dagger \chi)\bigg)
\end{split}
\end{equation}
Clearly, we can see that under the action of Galilean conformal symmetry generators, the variation in Lagrangian either changes by a total time derivative term or vanishes trivially. This assures that Galilean conformal symmetries are preserved at the level of \emph{action} in $d=4$ dimensions. The invariance of the Lagrangian under $J_{ij}$ is trivially satisfied (i.e, $\delta_J \mathbb{L}=0$). We can then conclude that Galilean Yukawa theory \eqref{eqn:complag} is invariant under infinite Galilean conformal symmetry generators.

\subsubsection{Conserved Charges}
Noether theorem suggests that, for every continuous symmetry of the Lagrangian, there exists a corresponding global conserved charge. Owing to the existence of infinite number of symmetries, we can deduce that Galilean Yukawa theory admits an infinite tower of conserved charges. The aim of this subsection is to construct those charges. We begin by briefly outlining the systematic procedure we would employ to construct the charges. Let us consider a generic Lagrangian $\mathbb{L}$ in $(d+1)$-spacetime which depends upon the general field $\Phi$ i.e,
\begin{equation}
\mathbb{L} \equiv \mathbb{L}(\Phi, \partial_t \Phi, \partial_i \Phi)
\end{equation}
Now consider the transformation of the field $\Phi \to \Phi +\delta_1\Phi$. If we invoke the Euler Lagrange equations of motion, the Lagrangian can at most change by a total time derivative i.e, we are studying the variation of the Lagrangian on-shell -
\begin{equation} 
\delta \mathbb{L} \bigg|_{\text{\tiny{on-shell}}}= \bigintssss d^d x\; \partial_t \Theta_t(\Phi, \partial_t \Phi, \partial_i \Phi, \delta_1\Phi)
\end{equation}
Now if we consider the infinitesimal symmetry transformation instead i.e, $\Phi \to \Phi+\delta_2 \Phi$ then the Lagrangian should differ only by a total derivative i.e,
\begin{equation} 
\delta \mathbb{L} \bigg|_{\text{\tiny{off-shell}}}= \bigintssss d^d x\; \partial_t \alpha_t(\Phi, \partial_t \Phi, \partial_i \Phi, \delta_2\Phi)
\end{equation}
Since, the symmetry transformations leaves the Lagrangian invariant i.e, choosing $\delta_1=\delta_2$ forces the off-shell variation to be equal to the on-shell variation,
\begin{equation*}
\delta \mathbb{L} \bigg|_{\text{\tiny{on-shell}}}=\;\;\;\; \delta \mathbb{L} \bigg|_{\text{\tiny{off-shell}}}
\end{equation*}
Thus we can deduce,
\begin{equation*}
\partial_t \big(\Theta_t -\alpha_t\big)=0
\end{equation*}
Hence, the corresponding global conserved charge is given by
\begin{equation}
Q=\bigintssss d^d x \;\big(\Theta_t -\alpha_t\big) 
\end{equation}
The Noether's procedure described above allows one to deduce the conserved charges associated to the Galilean conformal symmetry generators $\{L^{(n)}, M_{i}^{(n)}, J_{ij}\}$ in $(3+1)$--dimesnions. The on-shell variation of the Lagrangian \eqref{eqn:L} leads to
\begin{equation*}
\label{eqn:theta}
\Theta_t =-\sqrt{2}i \phi^\dagger \delta_1 \phi
\end{equation*}
Now, the off-shell variation of \eqref{eqn:L} under $L^{(n)}$ and $M_{i}^{(n)}$leads to 
\begin{eqnarray*}
&\alpha_t \bigg|_{L^{(n)}}&=f_n(t)\Bigg(\frac{1}{2}(\partial_i \varphi)( \partial_i \varphi )-\sqrt{2} i \phi^\dagger \partial_t \phi -\chi^\dagger \sigma^i \partial_i \phi \nonumber \\
&\qquad& \qquad \qquad +\phi^\dagger \sigma^i \partial_i \chi-g\varphi(\chi^\dagger \phi+\phi^\dagger \chi)\Bigg)\\
&\alpha_t \bigg|_{M_{i}^{(n)}}&=0\\
\end{eqnarray*}
where $f_n(t)=t^{n+1}$ is a Laurent polynomial in $t$.  The conserved charges for Galilean Yukawa theory becomes
\begin{align}
\begin{split}
\label{eqn:Ql}
Q_{L^{(n)}}=&\bigintsss d^3 x \; \Bigg\{ -\sqrt{2} i \dot{f}_n \phi^\dagger \bigg(x^j \partial_j \phi+\frac{3}{2} \phi\bigg)\\
&-f\bigg(\frac{1}{2}(\partial_i \varphi)( \partial_i \varphi )-\chi^\dagger \sigma^i \partial_i \phi +\phi^\dagger \sigma^i \partial_i \chi \\
&-g\varphi(\chi^\dagger \phi+\phi^\dagger \chi)\bigg)\Bigg\} 
\end{split}\\
\label{eqn:Qm}
Q_{M^{(n)}}=&-\bigintsss d^3 x \; \sqrt{2}i \phi^\dagger (\xi ^i \partial_i \phi)
\end{align}
where $\xi=\xi^j\partial_j$ is a spatially constant vector in time with $\xi^j =t^{(n+1)}(1,1,1)$. The finite conserved charges can be deduced from \eqref{eqn:Ql}-\eqref{eqn:Qm} by appropriately restricting $n$ to $(-1,0,1)$. For example, restricting $n=-1$ in \eqref{eqn:Ql} leads to the Noether charge corresponding to the time translations (Hamiltonian) i.e, 
\begin{equation}
\begin{split}
Q_{L^{(-1)}}=-\bigintsss d^3 x\; \bigg(\frac{1}{2}(\partial_i \varphi)( \partial_i \varphi )-\chi^\dagger \sigma^i \partial_i \phi \\
+\phi^\dagger \sigma^i \partial_i \chi-g\varphi(\chi^\dagger \phi+\phi^\dagger \chi))\bigg)
\end{split}
\end{equation}
which correctly reproduces the canonical Hamiltonian \eqref{eqn:hamiltonian}. This example also serves as a verification check for the conserved charges obtained for the Galilean Yukawa theory. In a similar fashion, we can construct other finite charges for Galilean Yuakwa theory by appropriately restricting $n$ to $-1,0$ or $1$ in \eqref{eqn:Ql} and \eqref{eqn:Qm}. A similar analysis yields the Noether charge for rotations $(J)$ i.e, 
\begin{equation}
Q_{\omega}= \bigintsss d^3x\; \bigg( -2 \sqrt{2} i \omega^{ij} x_i \phi^\dagger \partial_j \phi + \frac{1}{2} \omega^{ij} \epsilon_{ijk} \sigma_k \phi^\dagger \phi \bigg)
\end{equation}
where $\omega^{ij}$ is a constant antisymmetric matrix i.e, $\omega^{ij}=-\omega^{ji}$. 

\section{Quantum field description of Galilean Yukawa theory}
\label{section:quantum}
In the last section, we provided the classical field description of Galilean Yukawa theory. In this section we delve further into the theory by exploring the quantum field description of Galilean Yukawa theory. Our motivation to study the quantum `properties' stems from the realization that the theory admits infinite number of conserved charges (\eqref{eqn:Ql} and \eqref{eqn:Qm}) at classical level. It shall be interesting to understand how the symmetries behave at the quantum level. To this end, our primary goal is to construct a quantum field description of Galilean Yukawa theory. Our analysis relies upon the functional techniques. For the sake of brevity, we shall revert to \eqref{eqn:lagnonrel} to exploit the quantum nature of the theory. The \emph{action} $S$, for the theory reads,
\begin{equation}
\label{eqn:action}
S=\bigintsss dt \; d^3 x\;  \Bigg( \frac{1}{2}(\partial_i \varphi)( \partial_i \varphi )+i \bar{\psi}(\gamma^t \partial_t +\gamma^i \partial_i)\psi -g \bar{\psi}\psi \varphi\Bigg)
\end{equation}
The fermionic field $\psi$, carries a mass dimension $[\psi]=[\bar{\psi}]=3/2$ and the scalar field $\varphi$ admits $[\varphi]=1$. This restricts the coupling strength $g$ to dimensionless i.e, $[g]=0$ which makes for the case of a marginally renormalizable theory. For the rest of this paper, our focus will be on the 1 loop renormalization of Galilean Yukawa theory in $(3+1)$--dimensions. 

\subsection{Feynman Rules}
For notational convenience, we introduce  $\boldsymbol{p}=(\omega, p_i)$, where $\omega$ is the energy associated to the field and $p_i$ is the spatial momentum of the field. The Feynman rules for Galilean Yukawa theory \eqref{eqn:action} are
\begin{equation*}
\label{eqn:feynrule}
\begin{split}
&1. \qquad \text{For scalar propagator,} \qquad G(p, \omega)= \dfrac{i}{p^2} \\
&2. \qquad \text{For fermion propagator,} \;\; D(p, \omega)= \dfrac{i}{\gamma^t \omega+\gamma^i p_i} \\
&3. \qquad \text{For vetex,} \;\;\; \qquad V= -i g \qquad \qquad \qquad\\[4pt]
&4. \qquad \text{Overall multiplicative factor of -1 for each internal} \\
& \qquad \;\; \text{ fermion loop.}
\end{split}
\end{equation*}
\begin{table}
\begin{center}
\begin{tabular} { | m{1cm} | m{2cm} | m{4cm} | }
\hline
1. & $G(\omega,p_i)$ &\qquad \includegraphics[scale=0.6]{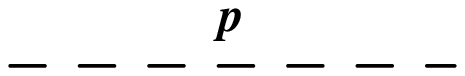}\qquad \;\\[7pt]
\hline
2. & $D(\omega,p_i)$ &\qquad \includegraphics[scale=0.6]{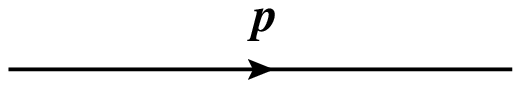}\qquad \;\\[7pt]
\hline
3. & $V$ &\qquad \includegraphics[scale=0.5]{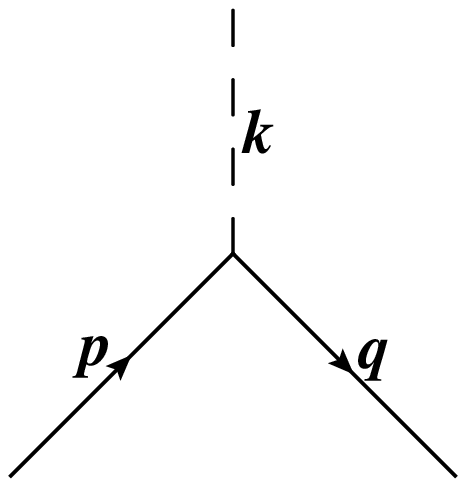}\qquad\\[7pt]
\hline
\end{tabular}
\caption{\label{fig:rule}Feynman rules for Galilean Yukawa Theory}
\end{center}
\end{table}
Diagrammatic representation of Feynman rules is given in \autoref{fig:rule}. Note that if we suppress the scalar degree of freedom, we end up with a pure Galilean fermion theory whose quantization is described by the fermion propagator. Tree level quantization (from both, canonical and functional methods) of free Galilean fermions has been studied in \cite{Santos:2005kk}\cite{Montigny2008PathintegralQO}. Also, if we suppress the fermionic part of the Lagrangian, we end up with a Galilean scalar field. An uninteresting feature of Galilean scalar field theory is that it does not admit any kinetic terms. Due to the lack of scalar dynamical degrees of freedom, we have not considered any self interaction term such as $\varphi^4$. Our point of interest lies in incorporating the matter degrees of freedom which happens to be fermionic fields $\psi$, in our case. In the next section, we study the 1 loop corrections to the propagators and vertex.
\subsection{Regularization}
\label{section:regular}
Owing to the existence of a three point vertex in the Feynman rules, the theory admits a correction to the scalar propagator, fermion propagator and vertex. In the general treatment of quantum field theory, the loop corrections are generally UV divergent quantities which must be regularized by restricting the infinite modes in momentum and energy integrals upto a UV cut-off regulator. In the context of Galilean field theories, the unequal footing of space and time forces one to consider the two cut-off regulators viz. $\Omega$ in the energy sector and $\Lambda$ in the momentum sector. Following along the lines of \cite{Banerjee:2022uqj}, we define the superficial degree of divergence by a set of two number $(\mathbb{D},\mathbb{F})$ i.e,
 \begin{eqnarray}
 \label{eqn:degreeD}
& \mathbb{D}&= 
\left(\begin{array}{c}
\text { Powers of $\omega$} \\
\text { in the numerator }
\end{array}\right)-\left(\begin{array}{c}
\text { Powers of $\omega$} \\
\text { in the denominator }
\end{array}\right)\\[7pt]
 \label{eqn:degreeF}
&\mathbb{F}&=\left(\begin{array}{c}
\text { Powers of $\vec{p}$} \\
\text { in the numerator }
\end{array}\right)-\left(\begin{array}{c}
\text { Powers of $\vec{p}$} \\
\text { in the denominator }
\end{array}\right)
 \end{eqnarray}\\
 The knowledge of superficial degree of divergence is helpful in understanding the extent to which the divergences can appear in the theory. However, as is often the case, the actual degree of divergence can be softer than the predicted superficial degree of divergence \cite{Peskin:1995ev}\cite{ryder_1996}. In what follows, we shall put \eqref{eqn:degreeD} and \eqref{eqn:degreeF} into use whenever required. We begin by evaluating the correction to the fermion propagator. 
 
 \subsubsection{Correction to the fermion propagator} 
\begin{figure}[h]
\centering
\includegraphics[scale=0.75]{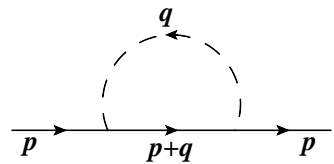}
\caption{Correction to the fermion propagator}
\label{fig:self1}
\end{figure}

 The Feynman diagram for the same is given by \autoref{fig:self1}.
\\
The loop integral $(\Sigma)$ corresponding to the figure \ref{fig:self1} can be evaluated by integrating along the unconstrained variables $(\omega_q, \vec{q})$ i.e,
\begin{equation}
\begin{split}
\Sigma(\omega_p, \vec{p})=\frac{i (-ig)^2}{(2 \pi)^4} \bigintsss &d \omega_q d^3 q \; \bigg(\frac{i}{\gamma^t(\omega_p+\omega_q)+\gamma^i(p_i+q_i)}\bigg) \\
& \times \frac{i}{q^j q_j}
\end{split}
\end{equation}
which can be rearranged to
\begin{equation}
\begin{split}
\Sigma(\omega_p, \vec{p}) = \frac{i (-ig)^2}{(2 \pi)^4} \bigintsss & d \omega_q d^3q\; \frac{1}{q^2 (p+q)^2}\\
& \times  \Big(\gamma^t(\omega_q+\omega_p)+(\gamma^i (p_i+q_i))\Big)
\end{split}
\end{equation}
The superficial degree of divergence is given by $(2,0)$, suggesting a quadratic divergence in the energy sector and a logarithmic divergence in the momentum sector. We introduce the cut-offs $\Omega$ in the energy sector and $\Lambda$ in the momentum sector. The integral evaluates to take the following value,
\begin{equation}
\label{eqn:div1}
\Sigma(\omega_p,\vec{p})= -i \frac{g^2}{8 \pi |\vec{p}|}\Big(\gamma^t \omega_p +\gamma^i p_i \Big) \Omega
\end{equation} 
Note that the loop integral above does not exhibit any logarithmic divergence offered due to the infinite modes of the momentum as predicted by the superficial degree of divergence. This is because the integral linear in $\vec{q}$ vanishes due to the anti-symmetricity of the integral. Also, the actual degree of divergence offered due to the energy sector does not agree with the superficial degree of divergence. This behaviour is not surprising since we know that the superficial degree of divergence renders only a naive idea about the extent of divergences in the theory. We note that in the cut-off limit $\Omega$, the integral $\int d\omega_q \gamma^t \omega_q$ vanishes owing to the fact that the integrand is an odd function of $\omega_q$. Hence, the true degree of divergence of the integral turns out to be linear. 

\subsubsection{Correction to the scalar propagator}
\begin{figure}[h]
\centering
\includegraphics[scale=0.75]{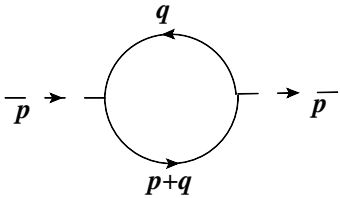}
\caption{Correction to the scalar propagator}
\label{fig:self2}
\end{figure}

The Feynman diagram for the correction to the scalar propagator is given in figure \ref{fig:self2}.
\\
The loop integral $(\Pi)$ in figure \ref{fig:self2} takes on the following form,
\begin{equation}
\begin{split}
\Pi (\omega_p, \vec{p})= \frac{-i}{(2 \pi)^4} \bigintsss d\omega_q d^3 q \; (-i g)^2 \; \text{Tr} \Bigg( \bigg( \frac{i}{\gamma^t \omega_q+\gamma^i q_i} \bigg ) \\
\bigg(\frac{i}{\gamma^t(\omega_q+\omega_p)+\gamma^j (q_j+p_j)}\bigg)\Bigg)
\end{split}
 \end{equation}
 The superficial degree of divergence in this case can be evaluated to $(1,1)$. This suggests that the loop integral diverges linearly with the momentum cut-off and energy cut-off respectively. As it turns out, the integal can be evaluated to the following value, 
 \begin{equation}
\label{eqn:div2}
 \Pi(\omega_p, \vec{p})= \frac{3 i g^2}{2 \pi^3} \Omega \Lambda
 \end{equation}
 We note that the 1 loop correction to the scalar propagator diverges linearly with both energy as well as momentum cut-off which is in agreement with the superficial degree of divergence. 
 
 \subsubsection{Correction to the vertex}
\begin{figure}[h]
\centering
\includegraphics[scale=0.75]{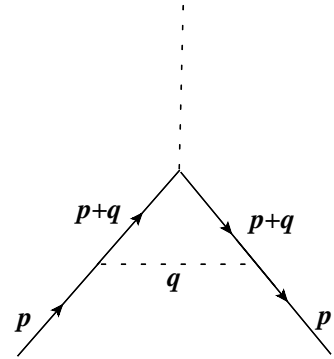}
\caption{Correction to the vertex}
\label{fig:self3}
\end{figure}

 The Feynman diagram for the correction to the vertex is given in figure \ref{fig:self3}. The loop integral $(\Gamma)$ takes the following form,
 \\
\begin{equation}
\begin{split}
\Gamma(\omega_p, \vec{p})=\frac{i}{(2 \pi)^4} \bigintsss d\omega_q d^3 q\; (-i g)^3 \frac{i}{q^j q_j} \\ 
\bigg( \frac{i}{\gamma^t(\omega_p+\omega_q)+\gamma^i (p_i+q_i)} \bigg)^2
\end{split}
\end{equation}
As before, we note that the superficial degree of divergence turns out $(1,-1)$. This is suggestive of linear divergence offered due to the infinite energy modes. Interestingly, the vertex correction in the momentum sector does not offer any divergence superficially. The integral can be evaluated to 
\begin{equation}
\label{eqn:div3}
\Gamma(\omega_p, \vec{p})= -\frac{i g^3}{8 \pi |\vec{p}|} \Omega
\end{equation}
Clearly, the integral diverges linearly.  We note that in all the three corrections, the integral necessarily diverges due to the cut-off offered at large energy values. This shall not be surprising at all given that the scalar propagator does not admit any kinetic term which means that we always have to integrate over infinite energy modes, rendering us a linear factor of $\Omega$ in each of the corrections.

\subsection{Renormalization and Beta function}
\label{section:renor}
\begin{figure}
\centering
\includegraphics[scale=0.5]{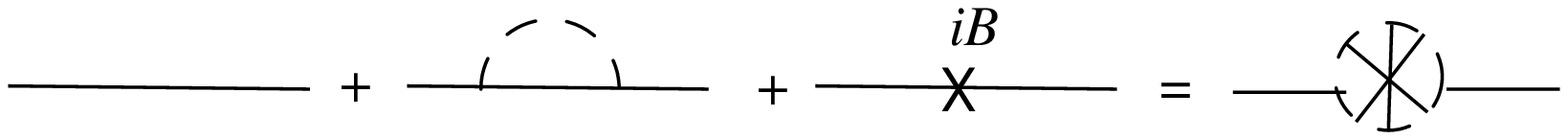}
\caption{Renormalization of fermion propagator}
\label{fig:counter2}
\end{figure}

Having evaluated all 1 loop corrections in the previous subsection, the stage is set to address the question of renormalization of the theory. As already mentioned, the dimensionless nature of the coupling strength $g$, makes Galilean Yukawa theory a reasonable candidate for a renormalizable theory. By introducing counter terms to the theory (i.e, subtracting the divergent pieces in the various cutoffs), we shall be able to absorb the divergent integrals \eqref{eqn:div1}, \eqref{eqn:div2} and \eqref{eqn:div3} in the field and coupling redefinitions.\\[5pt]
We begin our analysis with the correction to the scalar propagator. The correction to the scalar propagator \eqref{eqn:div2} suggests that we should add a following counter-term
\begin{figure}[h]
\centering
\includegraphics[scale=0.5]{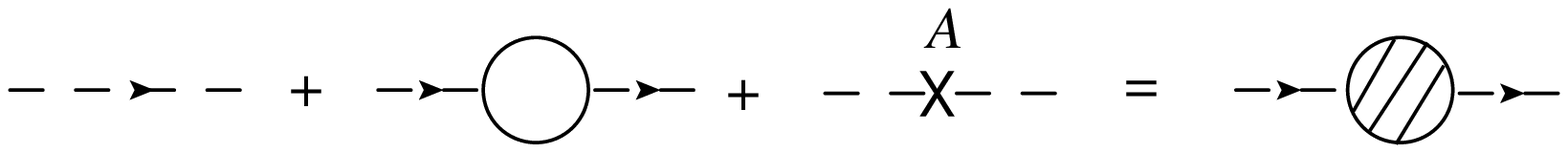}
\label{fig:counter1}
\end{figure}\\
where the coefficient $A$ can be chosen to render a finite propagator at 1 loop. The diagrammatic representation takes down the following mathematical expression 
\begin{equation*}
\begin{split}
&\frac{i}{p^2}+\frac{i}{p^2} \Bigg [ \frac{3 i g^2 \Omega \Lambda}{2 \pi^3}\Bigg ]\frac{i}{p^2}+\frac{i}{p^2} A\frac{i}{p^2} \;\; \equiv \;\text{finite}\\[5pt]
\implies & \frac{i}{p^2} \Bigg \{1+ \frac{i}{p^2}\Bigg[ \frac{3 i g^2 \Omega \Lambda}{2 \pi^3}+A\Bigg] \Bigg\} \qquad \equiv \; \text{finite}
\end{split}
\end{equation*}
After a bit of simple algebra, one can reduce the above expression to
\begin{equation*}
\dfrac{i}{p^2- i\Bigg[\dfrac{3 i g^2 \Omega \Lambda}{2 \pi^3}+A \Bigg]} \qquad \equiv \; \text{finite}
\end{equation*}
Clearly, for the above expression to yield a finite value, we must have the quantity in the bracket to vanish identically i.e,
\begin{equation*}
\dfrac{3 i g^2 \Omega \Lambda}{2 \pi^3}+A =0
\end{equation*}
which restricts $A$ to\footnote{Note that subtracting only the divergent pieces is one possible renormalization scheme out of many. We could also add to $A$ a finite term, without spoiling the finiteness of $A + 3ig^2\Omega\Lambda/(2\pi^3)$. This ambiguity is fixed by measurement, according to the standard renormalization procedure. For more details on renormalization see \cite{ryder_1996}.}
\begin{equation}
A=-\dfrac{3 i g^2 \Omega \Lambda}{2 \pi^3} = -i m^2
\end{equation}
where $m^2=\dfrac{3 g^2 \Omega \Lambda}{2 \pi^3}$ is the \emph{mass parameter}. Evidently, the propagator renormalization of the scalar field has introduced a \emph{mass} scale in our theory. The corresponding counter term in the Lagrangian is 
\begin{equation}
\label{eqn:ct1}
(\mathcal{L}_1)_{ct}= -\dfrac{1}{2} m^2 \varphi^2
\end{equation}
Remarkably, the scalar field has acquired the \emph{mass} under renormalization. The emergence of \emph{mass} term signals the breaking of conformal feature of Galilean Yukawa theory. However, the interesting thing to note is that $\varphi$ is a non dynamical field whose renormalization results in the \emph{mass} scale at quantum level. The emergence of a mass term for a non dynamical field is something that has never been seen in Galilean field theories. 
Note that the appearance of the mass term under renormalization is not surprising. In fact there is nothing sacrosanct about the emergence of mass in a renormalized theory. We shall recall that in Lorentzian massless $\varphi^4$ theory, the propagator correction leads to the mass term\cite{ryder_1996}. \\[5pt]
Now let us turn our attention towards the renormalization of fermion propagator. Diagrammatically, we can represent this in figure \ref{fig:counter2}.
Following along the lines of scalar propagator renormalization, we can deduce that the counter term takes the following form
\begin{equation}
(\mathcal{L}_2)_{ct}=i B \bar{\psi}(\gamma^t \partial_t+\gamma^i \partial_i)\psi
\end{equation}
where the coefficient $B$ should be fixed to absorb the divergences in the theory. Using (\refeq{eqn:div1}), we can fix $B$ as 
\begin{equation}
B=\frac{g^2}{8 \pi |p|} \Omega
\end{equation}
Our last hunt is to renormalize the vertex term. The counter term required to absorb the divergences in the vertex can be shown to take the following form
\begin{figure}[h]
\centering
\includegraphics[scale=0.5]{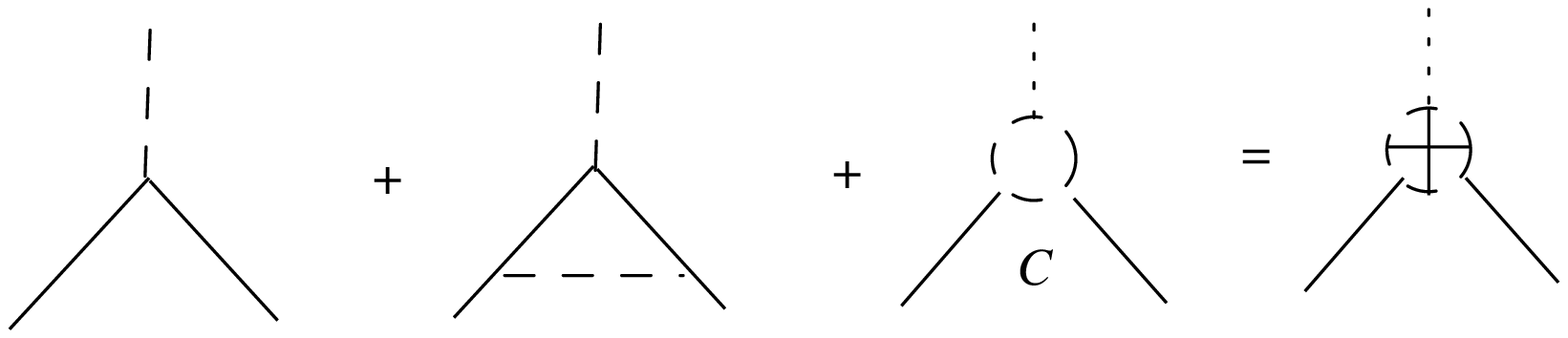}
\label{fig:counter3}
\end{figure}\\
where the coefficient $C$ can be fixed using \eqref{eqn:div3} to 
\begin{equation}
C=\frac{g^2}{8 \pi |p|} \Omega
\end{equation}
The counter term for the vertex in the Lagrangian takes the following form then
\begin{equation}
(\mathcal{L}_3)_{ct}= g C \bar{\psi}\psi \varphi
\end{equation}
Now that we have added the counter terms to our theory, we can define the bare Lagrangian $\mathcal{L}_b$, as 
\begin{equation*}
\label{eqn:barelag}
\mathcal{L}_b= \mathcal{L}+(\mathcal{L}_1)_{ct}+(\mathcal{L}_2)_{ct}+(\mathcal{L}_3)_{ct}
\end{equation*}
where $\mathcal{L}$ is given by \eqref{eqn:lagnonrel}. The bare Lagrangian takes the following form
\begin{equation*}
\begin{split}
\mathcal{L}_b&\;=\frac{1}{2}(\partial_i \varphi)( \partial_i \varphi )+i \bar{\psi}(\gamma^t \partial_t +\gamma^i \partial_i)\psi -g \bar{\psi}\psi \varphi \\
& -\frac{1}{2} m^2 \varphi^2+i B \bar{\psi}(\gamma^t \partial_t +\gamma^i \partial_i)\psi+Cg \bar{\psi}\psi \varphi
\end{split}
\end{equation*}
We can collect the coefficients to write
\begin{equation}
\begin{split}
\mathcal{L}_b=\frac{1}{2}(\partial_i \varphi)( \partial_i \varphi )-\frac{1}{2} m^2 \varphi^2 +(1+B)i \bar{\psi}(\gamma^t \partial_t +\gamma^i \partial_i)\psi \\
-(1-C)g \bar{\psi}\psi \varphi
\end{split}
\end{equation} 
We can then make the following redefinition for the fields
\begin{equation}
\begin{split}
\label{eqn:redef}
\varphi_{(b)}&=\varphi \\
\psi_{(b)}&=\sqrt{1+B} \; \psi
\end{split}
\end{equation}
The index $b$, appearing on the right hand side of the fields represent the bare field. Note that under renormalization, the scalar field does not get renormalized but instead leads to the \emph{mass} term in the theory. Using \eqref{eqn:redef}, we can write the bare Lagrangian as
\begin{equation}
\begin{split}
\label{eqn:bare1}
\mathcal{L}_b=\frac{1}{2}(\partial_i \varphi_{(b)})( \partial_i \varphi_{(b)} )-\frac{1}{2} m^2 \varphi^2_{(b)} +i \bar{\psi}_{(b)}(\gamma^t \partial_t +\gamma^i \partial_i)\psi_{(b)} \\
-\frac{g (1-C)}{(1+B)} \bar{\psi}_{(b)}\psi_{(b)} \varphi_{(b)}
\end{split}
\end{equation}
We can clearly see that with the choice \eqref{eqn:redef}, one must redefine the coupling as well i.e, 
Define,
\begin{equation}
\label{eqn:runcoup}
g_{(b)}= g \frac{(1-C)}{(1+B)}
\end{equation}
Thus, the bare Lagrangian (\refeq{eqn:bare1}) can be written down in terms of bare variables $\psi_{(b)}, \varphi_{(b)}$ and $g_{(b)}$ as,
\begin{equation}
\begin{split}
\mathcal{L}_b=\frac{1}{2}(\partial_i \varphi_{(b)})( \partial_i \varphi_{(b)} )-\frac{1}{2} m^2 \varphi^2_{(b)}+i \bar{\psi}_{(b)}(\gamma^t \partial_t +\gamma^i \partial_i)\psi_{(b)} \\
-g_{(b)}\bar{\psi}_{(b)}\psi_{(b)} \varphi_{(b)}
\end{split}
\end{equation} 
We note that even though, the theory can be made renormalizable, the bare Lagrangian does not share the same form as the starting Lagrangian. The emergence of the mass term for the scalar field is the captivating feature of Galilean Yukawa theory. As already explained before, the emergence of mass term is the signature of anomalous breaking of conformal symmetry. In order to see this explicitly, we shall construct the beta function $(\beta(g))$ for the theory. The significance of beta function is not just limited to conformal breaking of symmetry but in fact, it also helps us to understand the validity of the theory with the cut-off scale involved. The beta function is defined as 
\begin{equation}
\label{eqn:betafun}
\beta= \Omega \frac{\partial g}{\partial \Omega}
\end{equation}
We shall see from \eqref{eqn:runcoup} that 
\begin{equation}
g_{(b)}=g  \frac{\Big(1-\dfrac{g^2 \Omega}{8 \pi |p|}\Big)}{\Big(1+\dfrac{g^2 \Omega}{8 \pi |p|}\Big)}
\end{equation}
Since the running coupling $g$ is generally assumed to be small, we can condense the above expression to a simpler form by retaining the terms only upto $\mathcal{O}(g^3)$ i.e,
\begin{equation}
g_{(b)}=\Bigg (1-\frac{2 g^2 \Omega}{8 \pi |p|} \Bigg) g
\end{equation}
Since the bare coupling $g_{(b)}$ has to be taken independent of the cut-off, we can differentiate the above expression to arrive at
\begin{equation*}
\frac{\partial g}{\partial\Omega}= \frac{2 g^3}{8 \pi |p|} \Bigg(1+ \frac{6 g^2 \Omega}{8 \pi |p|} \Bigg)
\end{equation*}
Retaining the terms only upto $\mathcal{O}(g^3)$, we can write
\begin{equation}
\label{eqn:betatemp}
\frac{\partial g}{\partial\Omega}=\frac{2 g^3}{8 \pi |p|}
\end{equation}
Now note that in quantum field theories, the cut-off is often taken to be of the order of incoming momentum (or energy). Thus we can always define $\Lambda=b|p|$, where $b>0$ is a constant parameter. Also, the two cut-offs $\Omega$ and $\Lambda$ can be algebraically related by $\Omega=a \Lambda$, where $a\ne1$ is a positive constant that parameterize the discrepancy in the two cut-offs. These two conditions allow us to relate $\Omega$ with the momentum $|p|$ i.e,  in the limit $\Omega \to \infty$, $\dfrac{\Omega}{|p|}= ab$. We can then write \eqref{eqn:betatemp} as 
\begin{equation}
\label{eqn:asymbeta}
\beta(g)=\Omega \frac{\partial g}{\partial\Omega}= g^3 \bigg(\frac{ab}{4\pi}\bigg)
\end{equation}
Since we know that both $a$ and $b$ are strictly positive, this suggests that $\beta(g)$ is always positive i.e $\beta(g)>0$. This again confirms the presence of conformal anomalies in the theory and is in agreement with \cite{Jensen:2014hqa}\cite{Jain:2015jla}. We shall also make note of the fact that the theory is devoid of asymptotic freedom i.e, Galilean Yukawa theory becomes strongly coupled at large momentum (or energies). This becomes evident if we integrate \eqref{eqn:asymbeta} between a reference scale $\Omega_0$ and $\Omega$ i.e, we get, 
\begin{equation}
\label{eqn:solbeta}
g^2(\Omega)=\frac{g^2(\Omega_0)}{1-\dfrac{ab}{2\pi}g^2(\Omega_0)\ln \Big(\dfrac{\Omega}{\Omega_0}\Big)}
\end{equation}
It is straightforward to see from \eqref{eqn:solbeta} that at small values of momentum i.e $\Omega \sim \Omega_0$, we have $g(\Omega)\sim g(\Omega_0)$. However, at large momentum values i.e $\Omega >>\Omega_0$, the running coupling increases with the cut-off $\Omega$ confirming the invalidity of the theory at large energies. Another interesting thing to note here is the existence of Landau pole in the theory. We can check that \eqref{eqn:solbeta} shoots up at $\Omega=\Omega_0 \;exp\big(\frac{2 \pi}{a b \;g^2(\Omega_0)}\big)$. The existence of Landau pole is a feature often observed in quantum field theories that are not asymptotically free.
\section{Summary and Outlook}
\label{section:conclusion}
Let us summarize what we have accomplished in this paper. We have presented the classical and quantum field description of an interacting Galilean conformal field theory. We have taken the case of massless Galilean fermions coupled to a massless Galilean scalar field. The introduction of scalar-fermionic interaction incorporates the dynamical degrees of freedom into the free Galilean scalar field theory, which otherwise is an example of a non relativistic conformal field theory with non dynamical degree of freedom.  At the classical level, the Lagrangian for the theory is obtained by null reducing the Lagrangian for the relativistic Yukawa theory in one higher dimension. The resulting theory is found to be invariant under the full Galilean conformal algebra, hence the name Galilean Yukawa theory. We further exploit the presence of infinite symmetries in the theory by constructing the conserved charges \eqref{eqn:Ql}-\eqref{eqn:Qm} for the theory. The coupling strength in the theory is observed to be dimensionless which makes for the case of a marginally renormalizable theory i.e, the theory may or may not be renormalizable. Interestingly, what we have found is that the theory is renormalizable at least to 1 loop. Our prescription for quantization of Galilean Yukawa theory relies on path integral techniques. We regularize the UV divergences in the theory by setting the energy and momentum cut-off $(\Omega, \Lambda)$. An interesting feature that emerges out at quantum level is the entry of \emph{mass} in the scalar sector of the theory. The admission of \emph{mass} term in the Lagrangian suggests that the conformal invariance of the theory is broken at quantum level. This is captured by the behaviour of beta function which increase monotonically (and grows cubically) with the coupling. This suggests that the theory is not asymptotically free. The lack of asymptotic freedom is also captured by the Landau pole in the theory. Galilean Yukawa theory shares this interesting feature of anomalous breaking of conformal symmetry with Galilean quantum electrodynamics \cite{Banerjee:2022uqj}. However, an underlining difference between the two theory is the \emph{mass} term. It must be noted that Galilean quantum electrodynamics does not lead to any mass term. This is because the Galilean quantum electrodynamics is obtained as a null reduction of Lorentzian quantum electrodynamics in one higher dimension. It is well know that in relativistic setting, photon does not acquire mass under renormalization, courtesy of Ward identities. Thus the renormalization of gauge fields in Galilean limit is modelled in such a way that the gauge fields does not acquire mass. Galilean Yukawa theory is the first example of a Galilean field theory where \emph{mass} crops up. \\[5pt]
Note that the scalar field in this theory exhibit non dynamical degrees of freedom. The emergence of \emph{mass} term at the quantum level calls for further investigations since there is no precise notion of mass in Galilean setting. 
However, the theory share similar features to Galilean quantum electrodynamics. It must be noted that further studies on global conformal anomalies especially regarding anomalous Ward identities in the context of Galilean field theories might be more tractable with Galilean Yukawa theory than the gauge theory such as Galilean quantum electrodynamics. 
Also, that the quantum field description presented in this paper is only valid upto 1 loop in the perturbation. It shall also be very interesting to establish the renormalizability at all orders of the perturbations. \\[5pt]
Note that many of the recent studies \cite{Banerjee:2022uqj}\cite{Chapman:2020vtn}\cite{2022arXiv220706435B} on the quantization program of interacting Galilean field theories deal with matter-induced degrees of freedom. One of our future goal is to extend the quantization program developed in this paper to non abelian gauge theories such as Galilean Yang Mills (GYM)\cite{Bagchi:2015qcw}. GYM is an example of a self interacting theory hence it shall be interesting to explore the quantization of pure GYM in this setting. Also, studying the quantum ``properties" of Galilean QCD will be an avenue of future research.\\[5pt]
We also wish to extend the quantization program for Carrollian field theories ($c \to 0$, a degenerate twin of Galilean field theories) by developing a prescription similar to the one described in this paper. Carrollian physics has recently gained attention much to the fact that it plays an essential role in understanding gravity in asymptotically flat spacetime \cite{Duval:2014uva}\cite{Hartong:2015usd}\cite{Cardona:2016ytk}. Carrollian theories are promising candidate to study flat space holography. Also, recent study carried out with Carroll fluid allows one to model Carroll fluid as possible dark energy candidate \cite{deBoer:2021jej}. It shall be interesting if we can manage to uncover some interesting physics by probing the quantum properties of Carrollian field theories developed (see for example \cite{Banerjee:2020qjj}\cite{Bagchi:2019xfx}\cite{Bagchi:2022owq}\cite{Islam:2023rnc} and references therein) in recent years. Our recent work on renormalization of scalar Carrollian electrodynamics \cite{mehra2023towards} is a step in this direction. 

\section*{Acknowledgement}
AS would like to thank Kinjal Banerjee and Peter Horvathy for useful discussion on Galilean fermions. AS would also like to thank Stefano Baiguera, Lorenzo Cederle and Silvia Penati for several discussions on renormalization of non relativistic field theories. A note of thanks goes to JaxoDraw for developing their free Java program \cite{Binosi:2003yf} for drawing Feynman diagram.

\bibliographystyle{unsrt}
\bibliography{ref}

\end{document}